# A Test-bed Implementation for Securing OLSR in Mobile Ad-hoc Networks

Emmanouil A. Panaousis, George Drew, Grant P. Millar, Tipu A. Ramrekha and Christos Politis

*Wireless Multimedia & Networking (WMN) Research Group*
*Kingston University London, United Kingdom*

## ABSTRACT

*Contemporary personal computing devices are increasingly required to be portable and mobile enabling user's wireless access, to wired network infrastructures and services. This approach to mobile computing and communication is only appropriate in situations where a coherent infrastructure is available. There are many situations where these requirements are not fulfilled such as; developing nations, rural areas, natural disasters, and military conflicts to name but a few. A practical solution is to use mobile devices interconnected via a wireless medium to form a network, known as a Mobile Ad-hoc Network (MANET), and provide the services normally found in wired networks. Security in MANETs is an issue of paramount importance due to the wireless nature of the communication links. Additionally due to the lack of central administration security issues are different from conventional networks. For the purposes of this article we have used the "WMN test-bed" to enable secure routing in MANETs. The use of cryptography is an efficient proven way of securing data in communications, but some cryptographic algorithms are not as efficient as others and require more processing power, which is detrimental to MANETs. In this article we have assessed different cryptographic approaches to securing the OLSR (Optimised Link State Routing) protocol to provide a basis for research. We conclude the paper with a series of performance evaluation results regarding different cryptographic and hashing schemes. Our findings clearly show that the most efficient combination of algorithms used for authentication and encryption are SHA-1 (Secure Hash Algorithm-1) and AES (Advanced Encryption Standard) respectively. Using this combination over their counterparts will lead to a considerable reduction in processing time and delay on the network, creating an efficient transaction moving towards satisfying resource constraints and security requirements.*

## 1. INTRODUCTION

Mobile Ad-hoc NETworks (MANETs) are a grouping of mobile devices connected wirelessly in an ad-hoc fashion to form a coherent network structure enabling devices not directly connected and geographically separated to communicate and share resources using multi-hop routing. The applications of MANETs are ideally suited to situations where a coherent network infrastructure is unavailable or nonexistent, such as for military use in the field of operations or for use by medical staff in third world countries where the infrastructure does not exist. Another important application for MANETs is in emergency situations [1] such as terrorist attacks where the infrastructure is unavailable.

Due to the dynamic nature of MANETs with nodes joining and leaving frequently, routing protocols are essential to maintain an up-to-date picture of the MANETS topology. The MANET routing protocols enable nodes to discover routes to nodes they wish to communicate with by maintaining information regarding other nodes in the network. Common protocols used in wired networks are inefficient for MANETs so dedicated protocols have been developed. Primarily two types of routing protocols are used, *proactive* and *reactive*. Proactive protocols such as the Optimized Link State Routing (OLSR) [2], [3] protocol, proactively maintain routes between nodes and route information by propagating route updates thorough the network. In contrast, reactive routing protocols only institute routes on request, an example of this architecture is the Ad-hoc On-demand Distance Vector (AODV) [4], the Dynamic source routing (DSR) [4] and the Dynamic MANET On-demand (DYMO) [5] protocols.

In this paper taking advantage of the strength of Security Architecture for the Internet Protocol (IPSec) [7] we have secured the MANET routing protocol using our test-bed named WMN. We have actually used a hybrid version of the IPSec protocol, which includes both *Authentication Header* (AH) and *Encapsulating Security Payload* (ESP) modes, as proposed in [8], to provide a *green solution* in regards to the energy consumption security solution for MANETs. These two modes of IPSec guarantee integrity, authentication and confidentiality for the MANET communication links. More precisely, authentication and integrity are satisfied by the AH protocol that utilises a hash algorithm along with a symmetric Advanced Encryption Standard (AES) [11] key to produce a Hash Message Authentication Code (HMAC). For the ESP protocol we have used 128-bit symmetric keys because AES is one of the fastest and cryptographically strongest algorithms. Based on research published in [9], only the transport mode of the IPSec protocol has been used in our test-bed since it has been proven appropriate for MANETs.

The key objectives of this paper are the following:
- Discussion of the most important issues about security in MANETs.
- Discussion of the most crucial security benefits of the hybrid model of IPSec in MANETs.
- Description of the *WMN test-bed* software and certainly of the security implementation for MANETs.
- Evaluation of the *WMN test-bed* results in terms of overhead that each mechanism introduces to MANETs.

## 2. BACKGROUND

### 2.1. Related Work

The vast majority of the papers in the literature examine the issue of securing OLSR either in a theoretic basis or using simulators such as the network simulator ns-2. A limited number of papers has been experimenting using a test-bed. Although in our work we have secured OLSR using a test-bed environment, we will also briefly discuss some previous published related work including solutions that highlight simulation results or theoretical descriptions of some secure extensions for OLSR. The novelty of our work is the use of a real time test-bed to evaluate the performance of secure OLSR in MANETs. Our mechanism allows secure video transmission over MANET links introducing affordable overhead, as we will show in the performance evaluation section. According to our knowledge this is the first paper implementing a secure OLSR version and analytically measure its performance using video transmission.

### 2.1.1 Related Theoretical Work

In [20] authors propose a secure version of OLSR that protects packets using identity-based cryptography and periodically or when necessary refreshes cryptographic keys using threshold cryptography. The protocol allows only non-malicious nodes to participate in the bootstrap process while it introduces improvements in routing setup and maintenance.

The paper [21] expresses using a formal language the different types of trust relations between nodes running OLSR. The authors present a formal textual description of the trust issues for OLSR that enable an effective interpretation of attacks against OLSR in terms of trust classes and relations. In this way they claim that they can set the conditions to use trust-based reasoning towards the mitigation of particular vulnerabilities of OLSR. For a more extended work on trust management issues for MANETs, [31] is a complete survey that the reader can refer to.

Furthermore, paper [22] proposes a security mechanism to be integrated into OLSR. This mechanism distributes asymmetric (public key) cryptographic keys between the nodes in the network and "global timestamps" are used to avoid replay attacks determining whether any message is "too old" or not. The strong assumption of this mechanism is that trusted nodes cannot be compromised.

In [19] authors present an overview of security attacks against OLSR version 2, called OLSRv2, and show that OLSRv2 provides some inherent protection whilst in [23] authors discuss their implementation of an extension of the OLSR source code appearing in [14]. Their solution is based on signing each routing control packet using a digital signature to authenticate the message. Another consideration of this implementation is a timestamp mechanism to avoid replay attacks.

Last but not least, the paper [24] proposes a mechanism to enhance the security of the OLSR against external attackers based on message signing and sender authentication. Authors also deal with the case in which an adversary compromises a trusted node. The mechanism is based on recording recent routing information such as HELLO messages and using this information to prove the link state of a node at a later time by a new ADVSIG control message.

**2.1.2 Simulation Based Related Work**

In this section we briefly mention some related work done within the realm of securing the OSLR protocol whilst the evaluation of the proposed mechanisms have been carried out using network simulator. The paper [25] proposes a new secure version of OLSR called *Security Aware Optimized Link State Routing* (SA-OLSR). The protocol does not need any specialised hardware (i.e. GPS) and complete information of the whole MANET whilst preventing many attacks. To validate SA-OLSR authors have implemented the protocol using the network simulator ns-2 simulating also a misrelay attack as a case study. They show that the attack can totally disrupt the operation of OLSR whilst SA-OLSR is not affected. The quantitative indication for the aforementioned observation is that SA-OLSR has higher packet delivery ratio than the OLSR in the presence of adversaries.

Moreover, in [26] authors propose a secure fully distributed algorithm for the OLSR based on the secret sharing idea. The algorithm is based on threshold cryptography and it has been implemented using the OPNET simulator. Simulation results show that the additional delay due to the security considerations is affordable and suitable to the OLSR routing specifications operating in a transparent way.

The paper [27] proposes a hybrid protection scheme for OLSR based on identity-based digital signatures and hash chains. Since only a part of the messages are signed the rest include an undisclosed value from the hash chain to enable lightweight authentication. In this manner adversaries can hardly insert additional and false routing messages even if the these are not signed. The protocol is implemented using ns-2 tools and the simulation results highlight the average measured channel utilization per second, for OLSR traffic for various network sizes, security overheads and signature to hash ratios̃

Within the realm of intrusion detection, the [28] implements an Intrusion Detection System (IDS) that runs in each MANET node. The IDS infers and detects possible attacks against OLSR by using a set of rules that locally check the integrity of the OLSR routing messages and the MRP behaviour. The authors have implemented this IDS using ns-2 and they have evaluated its performance in terms of false positive and false negative detection rates.

### 2.1.3 Test-bed Based Related Work

In [29] authors present a key management protocol, called SkiMPy, which allows MANET node to agree on a symmetric shared key, used in the beginning of the network's lifetime to exchange digital certificates. The same key can be used to provide data confidentiality too along with preinstalled certificates to provide node authentication with the need for a third trust party. SKiMPy has been developed as a plugin for the OLSR. Their evaluation results show that SKiMPy scales linearly with the number of nodes in worst-case scenarios.

The paper [30] proposes a distributed and self-organized security scheme for OLSR. The scheme is based on threshold cryptography mechanisms to ensure the integrity of the routing messages. In the performance evaluation results authors show that the delay introduced by the scheme is acceptable and suitable to the routing requirements.

According to our knowledge there is no other testb-bed work, which aims at securing OLSR for MANETs. On the other hand, the aforesaid papers [29] and [30]:

- do not measure the delay and the data traffic rate for video transmission over MANETs when the OLSR protocol is used
- do not thoroughly explain the counterparts of their test-bed leaving many questions answered to the reader.

In this paper we extensively describe our secure routing mechanism applied to OLSR protocol along with all the software tools and the steps that have been followed to install them creating a valuable material for any reader. Morever, proper explanations about the choise of our tools have been given and performance evaluation results have been illustrated to render the efficiency of the IPSec-based hybrid mechanism.

### 2.3. IPSec

The IPSec protocol suite is a group of cryptographic protocols, which are used to secure connections between hosts on an IP network. IPSec was standardised by the IETF, which set out how the protocols should be used and documented in order to enable interoperability between disparate systems. IPSec is designed to provide the following security characteristics: *authentication*, *non-repudiation*, *confidentiality* and *integrity*.

IPSec can achieve these security goals by creating Security Associations (SAs) between nodes. An SA contains the addresses of the participating nodes and the type of security to be used along with the algorithms that will be used in each instance. The SA also contains the keys, which will be used by algorithms. The keys differ in length depending on the type of algorithm used and must be unique. A policy is recorded in the Security Policy Database (SPD), which details how the SA is to be implemented. The policy specifies which mode (tunnel or transport) will be used, how, and when it will be used. There are two ways we can apply security to our IP packets using IPSec. By using an AH, or we can use an encapsulating security payload. Alternatively we can use both, each has a specific part to play in the security process and it is important to understand what each feature is used for.

The AH is used to provide authentication and integrity of the IP packet being sent between the nodes. It is not designed as a cryptographic function to provide confidentiality; this is performed by a separate mechanism. AH is simply used to confirm with whom we are communicating, preventing attacks such as; replay attacks and man-in-the-middle attacks. The AH achieves this by performing a hash operation (also known as message digest) on the entire packet (excluding fields that may be altered such as TTL) and adding the result known as an Integrity Check Value (ICV) to the packet. When the packet is received at the destination the AH is stripped off and a hash function performed, if the result of this hash equals the original value then we can be sure that the packet firstly, has not been tampered with and secondly is from whom we think. IPSec offers different hashing algorithms for use. The most popular are the MD5 (Message Digest-5) [17] algorithm and SHA-1 (Secure Hash Algorithm) [16], which have had extensive research and testing performed on them. AH does not just perform a standard hashing procedure, as this would provide little defence against an attack. Instead it uses a HMAC which is a secret value introduced when creating the ICV. Using an AH provides us with integrity and authentication but it will not provide confidentiality, for this we need to use ESP.

As previously discussed, confidentiality is where we prevent any unauthorised person accessing our data, and this is most commonly achieved using cryptography. IPSec uses ESP to secure the data in the packet by using cryptographic algorithms to encipher the data for transmission. The most commonly used algorithms are 3DES [9] and AES as these are widely accepted. IPSec offers two different ways to use ESP, *tunnel* and *transport* mode.

Transport mode is used to secure end-to-end communications between participating nodes. Transport mode can use AH or ESP, or both to create a secure IP connection. If only AH is to be used, then it performs its hashing operation and the packet is amended with the AH header placed after the IP header providing integrity. Note that if we wish to add confidentiality we must use ESP to secure the *data field*. Using tunnel mode provides the more commonly understood idea of the Virtual Private Network (VPN). Tunnel mode with authentication takes the original packet, adds the AH and then instead of using ESP to encrypt only the data field, the whole packet is encapsulated using ESP and a new IP header added at the beginning of the packet.

### 2.3.4 IPSec Advantages and Disadvantages

There are many pros and cons to using IPSec. Firstly an important advantage of IPSec is that open-source software implementations exist, which have been standardised so there are many free implementations out there such as *Openswan* [18]. An important issue is that as long as the standards are adhered to it is possible to adapt your implementation to accomplish what you require for your situation. IPSec is also modular, so adaptations or new implementations of authentication or encryption algorithms may be incorporated into the architecture with ease. It also integrates into the existing IP infrastructure seamlessly without the need to implement new hardware; this is also relevant to IPv6, which IPSec is compatible with. This reduces concerns should a network adapt to use IPv6 in the future.

Though IPSec is considered the best current standard for securing the communication of IP traffic, it is not considered necessarily a good example of how a security suite should be. Firstly there is no single reference of how to correctly use IPSec, instead information must be collected from different sources. IPSec was standardised by using a committee structure and as a result IPSec has attempted to encompass many areas in its application. This means that there are many different ways IPSec can be configured which can create issues with incompatibility which further lead to possible vulnerabilities due to poor configuration, interoperation, and implementation.

When data is sent across a network many factors conspire to slow it down. Imagine a node trying to send a message to another node, which is four-hops away in the network. The message, which is sent must be interrogated by each intermediate node in order for it to attain its destination. If you imagine that each intermediate node takes one second to interrogate and forward the packet, the time it takes the packet to travel can quickly build up. Delay is an important issue as this can be the difference between a file taking two minutes or two hours to download. It may be said that delay is not so much an issue anymore with the increasing bandwidth and faster processors available, but this is incorrect. The counter argument is that more data is being sent, larger files, are being transferred and data hungry applications are being utilized.

In regards to MANETs irrespective of the processing speeds and bandwidth available the more efficient the network can be the better. One must take into consideration that MANETs are a grouping of mobile devices, and as discussed these devices have finite resources, the major factor being power. Also the devices that participate in MANETs could be hetrogenous mobile devices with different hardware configurations, which means they have varying lengths of battery life. So it is clear that we need to be as efficient as possible in all situations. There are many capable security controls which could be applied to MANETs but if they are not efficient then they will possibly be more detrimental to the network than some security threats. What we are looking for is a green security solution to apply to the network layer of MANETs and we propose that we use a selection of viable security options and find out whether they are suitable for use in MANETs.

We propose to use IPSec to enable secure communication at the network layer. As previously discussed IPSec can satisfy the requirements of confidentiality, integrity, and authentication depending upon its configuration. We intend to use IPSec in transport mode, using AH and ESP to offer the security requirements. We are using the transport mode over tunnel mode, as it requires less processing. It is known that transport mode does not offer as high a level of security as tunnel mode but this is another example of the trade off. Using the *WMN test-bed* we have assessed a selection of different cryptographic algorithms used by the AH and ESP functions in IPSec to find which is the most efficient for use in MANETs. We have selected algorithms which are deemed secure (they have not been compromised as of yet) and they comprise of the following; *AH*: MD5 and SHA-1, *ESP*: AES and 3DES. In order to achieve this, a *WMN test-bed* scenario will be designed and implemented in order to realise the answers posed by the aforementioned questions. We intend to analyse and discuss the results in order to reach a conclusion and to suggest ways in which this may be used for future work.

## 3. WMN TEST-BED IMPLEMENTATION

### 3.1. Test-bed Setup

The WMN test-bed was utilised to find the overhead introduced by adding different cryptographic algorithms to a video streamed across a MANET. The requirements for setting up the test-bed are summarised in the following:

- Operating System: Each laptop was required to have Linux Ubuntu 9.10 Karmic Koala [12] installed.

- Wireless Interface: Each laptop was required to have a wireless card that could support IEEE 802.11g communications.

- Video Stream: In order to stream and receive Video each laptop was required to have VLC [13] player installed.

- Routing: For multi hop communication to be achieved, the OLSRd [14] routing protocol was installed and configured in all the laptops as described in [14].

- In order to provide secure communication using cryptographic algorithms the sending and receiving laptops needed to be configured with IPSec as described below.

- To record the communication among the devices, the WireShark [15] network analyser was installed in each node.

## 3.2. Testing Scenarios

### 3.2.1 Scenario I: Single Hop

The single hop scenario would be the first step in the process of testing and would provide a starting point for the testing to progress from, a set of results in order to compare it with the multi-hop scenario. The single hop scenario would require two laptops connected in an ad-hoc fashion and the scenarios would proceed as follows:
- Node *sender* transmits a UDP (User Datagram Protocol) video stream to node *receiver* using VLC Player.
- Node *sender* transmits encrypted with IPSec to node *receiver*.

Scenario 1.1 would provide the base line recording of communication without security. Scenario 1.2 would be repeated four times in order to gain result for a combination of authentication and encapsulating algorithms. These combinations are as follows: AES with MD5, 3DES with MD5, AES with SHA-1 and 3DES with SHA-1. This would provide a broad array of results for us to analyse against the base line of the UDP video stream so that the most efficient combination could be identified.

### 3.2.2 Scenario II: Multi Hop

This scenario would be more complicated than scenario I. It requires three nodes, *sender* node, *intermediate* node, and *receiver* node to be part of the MANET using the ad-hoc routing protocol OLSRd to maintain their routing tables in order to perform multi hop routing.

The procedure is similar to the first set of scenarios but requires multi hop routing to facilitate the forwarding of the packet through intermediate node to the intended receiver node as detailed below:
- The *sender* node streams UDP Video to *receiver* node via intermediate node.
- The *sender* node streams UDP Video encrypted with IPSec to node *receiver* via *intermediate*.
- As before, the first scenario 2.1 will provide a base line from which the other results may be compared against, and scenario 2.2 will provide the multiple security approaches we require.

## 3.3 Installation

This subsection describes the implementation process. This was achieved by first installing the required software. This subsection also contains details of any issues encountered during the installation process. For the test-bed three laptops were required and though this is not an ideal testing environment, real life comparisons can still be made. Even though standards laid down by standardisation bodies such as the IEEE improve compatibility these are by no means a guarantee that the guidelines have been followed correctly.

### 3.1.1. Operating System

The operating system (OS) used for testing is important and choosing the wrong OS can lead to compatibility issues. The OS that was used on the laptops was Linux Ubuntu 9.10 Kernel version 2.6; this selection is justified next. Linux is a UNIX based operating system developed under the GNU General Public Licence and is free. This means that there is no cost to provide a uniform stable operating system across all the devices in the test-bed. As Linux is open source software it is used widely and developers are welcome to offer any bug fixes as well as produce other stable programs enabling Linux to release solutions to common issues. Linux Ubuntu hardware requirements are less demanding than other operating systems such as Microsoft Windows or Mac OS this makes it a desirable OS for static as well as mobile devices. Additionally to this there are lightweight versions of Ubuntu for the net-book architecture. Moreover, the user is able to tweak the OS to his personal need, it is simple to download software packages you require and remove those that are adding unnecessary burden through unneeded processing.

### 3.1.2 VLC Player

It had been mentioned that the data that would be transmitted on the test-bed would be a UDP video stream. To enable this process the laptops required a media player capable of streaming and receiving video on a 802.11 network. One such media player is the VLC player. VLC is distributed as free open source software by the Video LAN project released under the *GNU General Public License*. The VLC is cross platform and there are versions for most common OS platforms such as Mac OS, Windows and Linux. Fortunately there is an explicit VLC version released for Ubuntu 9.10. VLC player was available from the *Synaptic Packet Manager* Linux GUI but could also be downloaded in the terminal. It is important to note that running the *update* command may affect other programs unless explicitly checked before running.

### 3.1.1. Configuring Ad-hoc Communication

Once Ubuntu had been installed successfully on all the laptops and the initial update process completed the next task was to configure each laptop so that they could communicate with each other in an ad-hoc fashion.

Node IP addresses were selected from the 192.168.2.0 network range. This is a private IP address range and is acceptable for use within a private network such as described above. Also, all the nodes are on the same network to keep the communication simple without introducing any cross network communication issues that can arise.

Once all nodes are configured correctly and have joined the MANET we may test that there is connectivity between the nodes using an ICMP echo message (PING). The reason for this is that there are many factors that can affect network communication as covered further in the test-bed section including issues such as firewalls or hardware compatibility with 802.11 standards although the wireless network adapters all appear on the Linux hardware compatibility list. For establishing node connectivity, it was ensured that all ICMP requests were replied to by all nodes meaning that in theory it should be possible to communicate in an ad-hoc fashion between the nodes.

### 3.1.1. Installation of OLSRd

The newest stable release available for download is *olsrd-0.5.6-r8* [14] (for Ubuntu OS) and this is the version used in this test-bed. The file olsrd-0.5.6-r8 is downloaded and extracted to a target file this can be done via the Linux GUI or from the terminal. OLSRd then needs to be compiled and installed, this is accomplished by navigating to the *olsrd-0.5.6-r8* directory in a terminal session and then issuing the commands `sudo make` then `sudo make install`.

When this process was executed it returned the same error on all of the laptops. The error detailed that *Flex* and *Bison* where missing and that these files needed to be installed before OLSRd could be correctly compiled and installed.

*Bison* is a program that generates parsers. A parser is a component part of a compiler and has many applications in different circumstances, but in general it is used as a text analyser, checking for the correct syntax of language and builds data structures. As this important program was missing, OLSRd would not compile correctly.

Then, *Flex* is a program used for generating scanners in OLSRd (as well as other programs). Flex is designed to recognise lexical patterns in text, it will read files to determine which

148

scanner to generate and create an output a source file to define the routine. The latter file is essential for the successful compiling of OLSRd. Both *Flex* and *Bison* were successfully installed issuing the commands `sudo apt-get install bison` and `sudo apt-get install flex`.

Now that the required scanner and parser generators were installed, the compile and install commands `sudo make` then `sudo make install` were issued again and the installation was successful. There were some additional setup requirements to be satisfied though we had to compile and install some plug-ins called *txtinfo* and *httpinfo*. This was accomplished by first navigating to the correct folders *downloads/olsrd-0.5.6-r8/lib/txtinfo* and *downloads/olsrd-0.5.6-r8/lib/httpinfo* and then entering the commands `sudo make` and then `sudo make install`.

### 3.1.3 Wireshark

When using WireShark with Linux platforms, some Kernel configurations are required so that WireShark is able to capture 802.11 packets correctly. When collecting 802.11 packets some interface devices are set to filter out certain packets such as management packets. The IEEE 802.11 packet headers are translated by the network driver into *fake* Ethernet headers. As a result, wireless interfaces have been designed to ignore traffic from other Service Set Identifiers (SSIDs) and to filter out as well as translate other data, implying that a true picture or *capture* in this case would not be attained. The main reason for using WireShark is to record data, that we can then analyze using various WireShark analysis tools for displaying data. The main tools used in this work were *Statistics Summary* and the *IO Graph*.

### 3.1.4 IPSec

There are various IPSec programs available for all manner of operating systems allowing interoperability, which is one of the aims of the IPSec suite.

**Cost:** Though IPSec is open source and only requires that the developers conform to the standardisation guidelines laid out by the IETF as previously discussed, means that there are many different approaches to IPSec. Although there are many expensive versions of IPSec, there exist may common free applications available to Linux users too. The most prevalent examples of IPSec packages are:

**Free Secure Wide Area Network (FreeS/WAN):** FreeS/WAN is a free implementation of IPSec and IKE for Linux. FreeS/WAN has been superseded by StrongS/WAN which incorporates improved IKE, NAT traversal and other functionality. Though this is a free and well-documented IPSec implementation there is extensive configuration needed.

**KAME-tools:** This is the *native* IPSec stack found in Linux Kernels 2.5.47 and 2.6 onwards.

For our implementation it is only required for two nodes to be configured with IPSec. This is because only the node streaming the video and the node receiving the video will need to participate in secure communication. Therefore there is no need for the intermediate node, in this case node B (192.168.2.2) to be configured with IPSec.

Firstly it is important to make sure that the correct kernel version is in place i.e. the 2.5.47 and 2.6 versions onwards that are available at http://www.kernel.org. These must be correctly extracted and compiled. The kernels installed on all the laptops were version 2.6* which satisfied the requirements. Next it was necessary to configure the kernel with the functions needed for IPSec.

Changes to the kernel are made using the `sudo make menuconfig` command, which executes a convenient GUI similar to a BOIS GUI, allowing functions to be switched on or off and then saved. The following settings were required to be made to the Kernel to allow the correct IPSec functionality. Depending on the version of the Kernel being used IPv6 support may need to be turned on.

The required libraries *ncurses* was missing. *ncurses* is a programming library that provides an API allowing a programmer/user to make changes to scripts. To add this missing library, we have attempted to download the *ncurses* file by running `sudo apt-get install ncurses` but this returned an error informing that the target no longer existed. After researching we found that the *ncurses* library had been updated and the correct file needed was in fact *ncurses-devel*. Once this library had been installed, the `sudo make menuconfig` command worked and kernel configuration was completed and saved.

```
Networking support (NET) [Y/n/?] y
  *
  * Networking options
  *
  PF_KEY sockets (NET_KEY) [Y/n/m/?] y
  IP: AH transformation (INET_AH) [Y/n/m/?] y
  IP: ESP transformation (INET_ESP) [Y/n/m/?] y
  IP: IPsec user configuration interface (XFRM_USER) [Y/n/m/?] y

Cryptographic API (CRYPTO) [Y/n/?] y
  HMAC support (CRYPTO_HMAC) [Y/n/?] y
  Null algorithms (CRYPTO_NULL) [Y/n/m/?] y
  MD5 digest algorithm (CRYPTO_MD5) [Y/n/m/?] y
  SHA1 digest algorithm (CRYPTO_SHA1) [Y/n/m/?] y
  DES and Triple DES EDE cipher algorithms (CRYPTO_DES) [Y/n/m/?] y
  AES cipher algorithms (CRYPTO_AES) [Y/n/m/?] y
```

Figure 1. IPSec Kernel Configuration.

The final step of installation was to download *ipsec-tools* that contain user space tools. These tools, as listed next, are used to configure and run integral parts of the IPSec implementation:

- *Libipsec*: Library with PF_KEY implementation.
- *setkey*: Tool to manipulate and dump the kernel *Security Policy Database* (SPD) and *Security Association Database* (SAD).
- *racoon*: IKE daemon for automatically keying IPSec connections.
- *racoonctl*: A shell-based control tool for racoon.

When configuring IPSec there are some choices to be made on the type of security one wishes to use. We intend to implement IPSec in *transport mode* with different types of cryptographic algorithms for AH (MD5/SHA-1) and ESP (3DES/AES).

For secure communication between two nodes using IPSec certain parameters need to be configured. A configuration file needs to be created in the *etc* directory called *setkey.conf* using a file editor such as *gedit* with the `sudo gedit /etc/setkey.conf` command. In **Figure 2**, such a setkey.conf file is shown where the configuration is for the 192.168.2.12 node to communicate with the 192.168.2.22. The first two flush commands are used to flush the *SAD* and the *SPD* respectively. This makes sure no previous configurations are left that may cause errors.

The encryption keys have been manually inputted into configuration files. The rationale for this approach is that as this is a test-bed containing a small amount of nodes the configuration and administration is low compared to many nodes. Though it is possible to use racoon and IKE to automatically distribute the correct keys this will add to the overhead on the network and requires considerable configuration, which is not viable in regards to this project.

As noted in the previous paragraphs the AH and ESP have different length keys this is because different implementations of encryption algorithms require different sized keys, it is also important that these keys remain secret. Below are the bit requirements for the encryption/hashing algorithms used for testing: 160 Bit for SHA-1, 128 Bit for MD5, 192 Bit for 3DES and 192 Bit for AES.

```
#!/usr/sbin/setkey -f

# Configuration for 192.168.2.12 MD5 and AES

# Flush the SAD and SPD
flush;
spdflush;

# AH SAs
add 192.168.2.22 192.168.2.12 ah 0x200 -A hmac-md5
 0xce516b2abf2fa2e6ab952f0454f7ab11;
add 192.168.2.12 192.168.2.22 ah 0x300 -A hmac-md5
 0xc2357ddcb7d2eb510448e716afecd4f2;

# ESP SAs
add 192.168.2.22 192.168.2.12 esp 0x201 -E aes-cbc
 0xb05e9caf66242c383903c367699ca452d0e8fa41f7aeab1d;
add 192.168.2.12 192.168.2.22 esp 0x301 -E aes-cbc
 0xd7ffecd485b1410d6d600598c14728962e4096ff9bf5ea42;

# Security policies
spdadd 192.168.2.22 192.168.2.12 any -P in ipsec
            esp/transport//require
            ah/transport//require;

spdadd 192.168.2.12 192.168.2.22 any -P out ipsec
            esp/transport//require
            ah/transport//require;
```

Figure 2. Setkey Configuration File (setkey.conf).

It is important that the keys used cannot be guessed and should be random in nature. While it is difficult to truly produce random results, Linux has a program called *dd* capable of producing random bit streams using the *random* or *urandom* statement. Below is an example of a *dd* command, which will produce a 128 Bit random digit key.

```
dd if = /dev/random count=16 bs=1|xxd –ps
```

To change the size of the bit stream output the *count* factor must be changed. In the above command the count is 16, this is because 128/8=16. If we instead required a 192-bit key the *count* would be 24 (192/8=24). Once both nodes have been configured the configuration file must be loaded which will start IPSec communication the command used is `sudo setkey –f /etc/setkey.conf`. After this command has been issued the running configuration may be tested by issuing the following commands:

```
'Sudo setkey –D': tests the SAD
'Sudo setkey –PD': tests the SPD
```

We found out that communication could be blocked by iptables as found with the OLSRd configuration. In some instances it was required to configure iptables to accept ESP and AH traffic in and out of the communicating nodes and also forwarding on intermediate nodes. In the commands below -p 50 relates to AH and the commands would need to be added with -p 51 to allow ESP to pass.

     **Inbound:** `sudo iptables – A INPUT –p 50 –j ACCEPT`
     **Outbound:** `sudo iptables – A OUTPUT –p 50 –j ACCEPT`
    **Forwarding:** `sudo iptables – A FORWARD –p 50 –j ACCEPT`

As previously demonstrated for the OLSRd configuration and testing the Linux native firewall was disabled for testing purposes.

```
Chain INPUT (policy ACCEPT)
target     prot opt source               destination
ACCEPT     esp  --  anywhere             anywhere
ACCEPT     ah   --  anywhere             anywhere

Chain FORWARD (policy ACCEPT)
target     prot opt source               destination
ACCEPT     ah   --  anywhere             anywhere
ACCEPT     esp  --  anywhere             anywhere

Chain OUTPUT (policy ACCEPT)
target     prot opt source               destination
ACCEPT     ah   --  anywhere             anywhere
ACCEPT     esp  --  anywhere             anywhere
```

Figure 3. Sudo iptables –L output.

## 4. WMN TEST-BED RESULTS

### 4.1. How Results Were Gathered

To show the performance evaluation results we have:

1. firstly used tested communication between one-hop neighbours, nodes which have a direct connection to each other as they are geographically in range of each other.

2. then added a third node, and tested communication between two- hop neighbours or two nodes which are geographically separated, but share a connection to an intermediate node. The nodes can communicate amongst themselves using a routing protocol. By recording data from multi-hop and single-hop neighbours we would be able to make comparisons between the two cases.

3. taken readings of standard communication between nodes in each case.

4. repeated the communication but added different security algorithms. The standard UDP communication results should act as a baseline for which we can base the results of applying security.

In order to conduct analysis we needed results. In this section, we detail the manner in which we collected data from the WMN test-bed. We scrutinize the parameters applied to the recording process, evaluate and discuss the results that are presented using network analyser called WireShark to achieve this.

When recording the results we needed different scenarios to test. Detailed in the following are the scenarios used in the testing process; one-hop UDP, multi-hop UDP, one-hop 3DES/MD5, multi-hop 3DES/MD5, one-hop 3DES/SHA-1, multi-hop 3DES/SHA-1, one-hop AES/MD5, multi-hop AES/MD5, one-hop AES/SHA-1 and multi-hop AES/SHA-1.

We needed to set a uniform limit for the duration of each test; it was possible to set this to different values such as packet volume, bytes of data or time periods. When deciding on the length of period we took certain issues into account. We wished to have a period that was long enough to collect sample data. If that period was too long then the volume of data would be vast and difficult to collate and assess. Conversely, if the period was to small this may not record enough data or give a representative account of how communication behave. In essence once the connection is established enough stable data is recovered to provide accurate trending. It was decided a period of five minutes would be appropriate to accumulate the required data.

## 4.1. Total Packets Sent

### 4.1.1. Single Hop

The numbers of packets rises when the scheme of SHA-1 with 3DES has been used displaying more packets sent than any of the others schemes. The sending and receiving packet numbers are generally of similar numbers as shown in **Figure 4**.

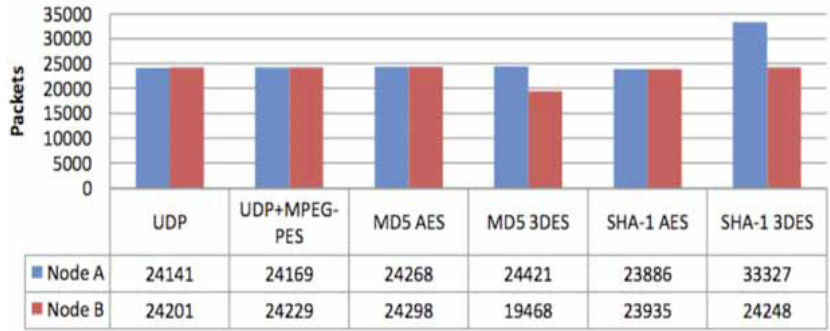

Figure 4. Total packets against different security schemes in a single hop scenario.

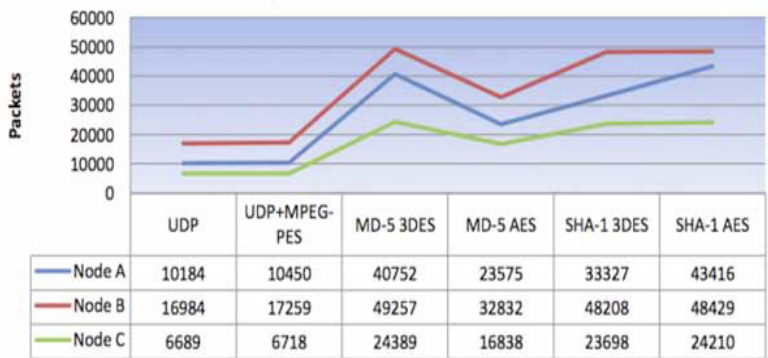

Figure 5. Total packets against different security schemes in a multi hop.

### 4.1.2. Multi Hop

It can be observed in **Figure 5** that there is a common trend in the results. The UDP results remain level but once the security is added in all cases there is a rise in the amount of packets being sent, forwarded and received. We can see in the results the rise in packets when 3DES is used and in general this protocol produces higher packet numbers. Although SHA-1 with AES is responsible for sending the most packets according to the results, these are not aligned with the general trends exhibited by the other protocols and this may be an anomaly.

### 4.1.3. Comparison

Placing these two sets of results in contrast it is clear that using security increases the number of packets sent on the network. Over one hop the rise is minimal but as another node is added the packet numbers increase dramatically. While analyzing the protocols we see that the AES combinations produce less packet traffic than the 3DES combinations.

## 4.2. Average Packet Size

### 4.2.1. Single Hop

The results for single hop, depicted in **Figure 6**, have been displayed as a bar chart. We notice that the results are the same with node A and node B mirroring each other. It can be seen that there is an average increase in packet size with the addition of the security. Also it can be noticed that AES is cheaper in terms of packet size than 3DES regardless of the authentication being used MD5 or SHA-1.

### 4.2.2. Multi Hop

The results in **Figure 7**, show a similar trend to the single hop results where the security introduces higher average packet sizes than the UDP base line, but with slight deviations in the UDP levels. The information from the multi hop data shows that the AES combinations have higher average packet sizes than 3DES.

### 4.2.3. Comparison

It seems confusing that 3DES would have higher packet sizes at single-hop and then AES has higher at multi hop. In fact, although this is noticeable through the results, there is only a relatively small difference between AES and 3DES at single hop, or multi hop. Hence, comparatively it is a matter of only 8 bytes overhead in most cases on an average. To pinpoint the situation it should be considered that AES has slightly larger average packet sizes than 3DES in a multi hop scenario.

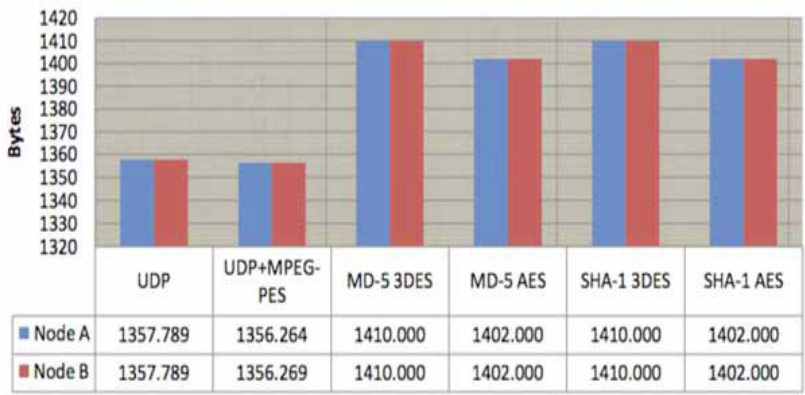

Figure 6. Average packet size against different security schemes in a single hop scenario.

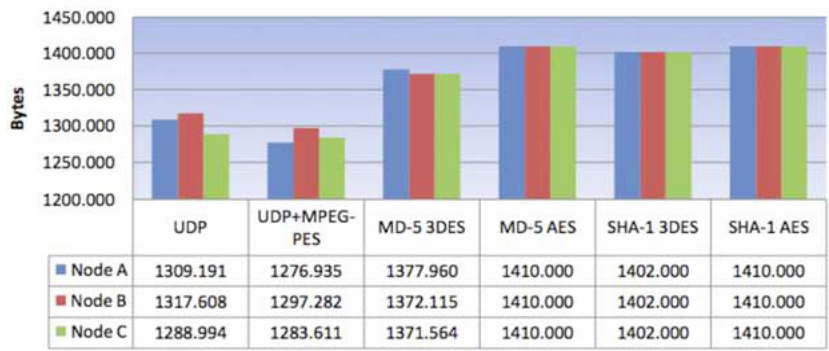

Figure 7. Average packet size against different security schemes in a multi hop.

International Journal of Network Security & Its Applications (IJNSA), Vol.2, No.4, October 2010

## 4.3. Bit Data Rate

### 4.3.1. Single Hop

In this case, as seen in **Figure 8**, there is a rise in the bit-rate when the security is added in the region of 0.025. There is a spike in the bit-rate by the sending node for SHA-1 with AES, showing that AES in general is producing higher bit rates than 3DES. As looking at the trends we would expect AES with SHA-1 to be equal on both nodes so it may be fair to surmise that a more realistic value of 0.909 is relevant here which still holds true the higher bit-rate of AES over 3DES.

### 4.3.2. Multi Hop

This is a set of interesting results as with the single hop results, some components do not completely follow the trend expected. For instance, the results for the sending node, as illustrated in **Figure 9**, displays an expected hike in bit-rates (they are higher than single hop) . With regards to the intermediate node and the receiving node the security protocols are quite similar and we would expect this from the receiving node as well. In any case the results show that AES has a higher bit-rate with regards to combining with SHA-1 and conversely 3DES has higher bit-rate over AES when combined with MD5.

### 4.3.3. Comparison

The comparison is difficult to make here because of the contrary results. At single-hop AES maintains higher bit-rates than 3DES, but part of this information does not translate to the multi hop scenario. Here the trend for the SHA- 1 combinations continue with AES continuing higher rates than 3DES, but when we scrutinize the MD5 combinations we see that on the whole 3DES has a higher bit-rate than AES. Taking this data into account the result we see is that multi-hop the highest bit-rates were achieved with AES/SHA- 1 and the lowest with AES/MD5.

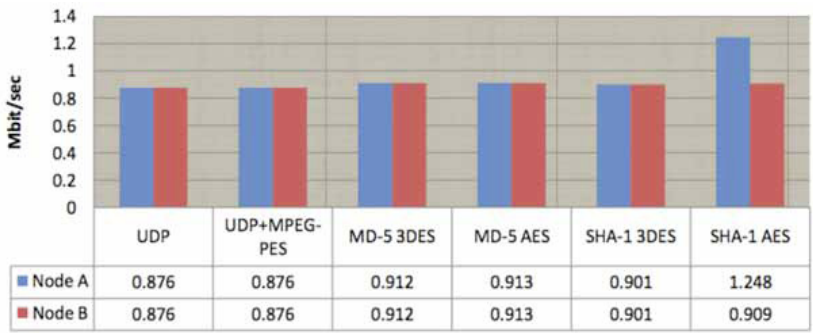

Figure 8. Bit data rate against different security schemes in a single hop scenario.

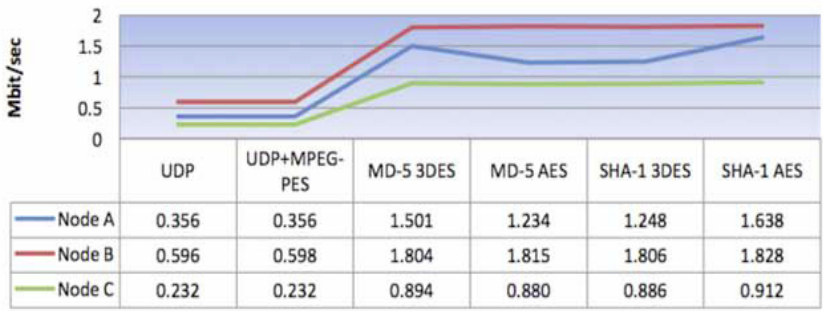

Figure 9. Bit data rate against different security schemes in a multi hop scenario.

## 4.4. Packet Data Rate

### 4.4.1. Single Hop

As with the results, depicted in **Figure 10**, there is a rise in value as we add the security protocols. Strangely, in the SHA-1 with 3DES scheme, data is actually less than the UDP base line data. This might be because of an issue with time periods, where the SHA-1/3DES time was slightly shorter than the UDP period due to an error in WireShark, which reduced the resulting average.

### 4.4.2. Multi Hop

In **Figure 11** is a clearer set of results which shows that to achieve the smallest amount of packets per second, then AES/MD5 should be used with 3DES/MD5 being the most costly. On average AES combinations had lower packets per second overall.

### 4.4.3. Comparisons

Due to the confusing nature of the single- hop results it is difficult to make a solid comparison between the two scenarios and the protocol combinations. It would be prudent to retake the results to assess if the same values appear. Given the results, we must make the cautious assumption that AES is preferred over 3DES in the multi-hop environment.

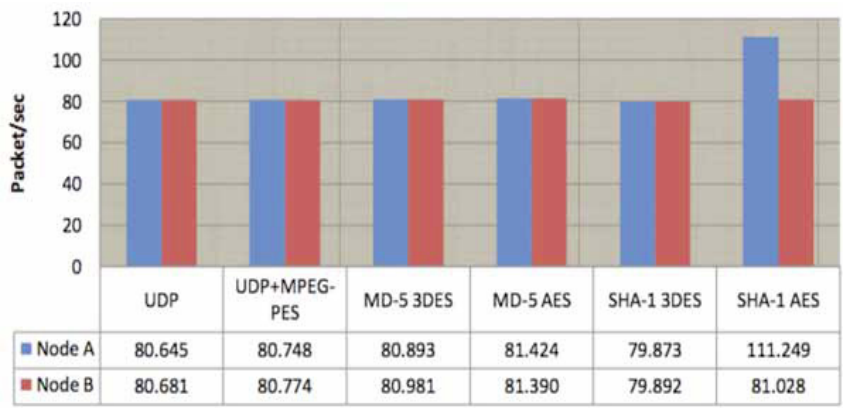

Figure 10. Packet data rate against different security schemes in a single hop scenario.

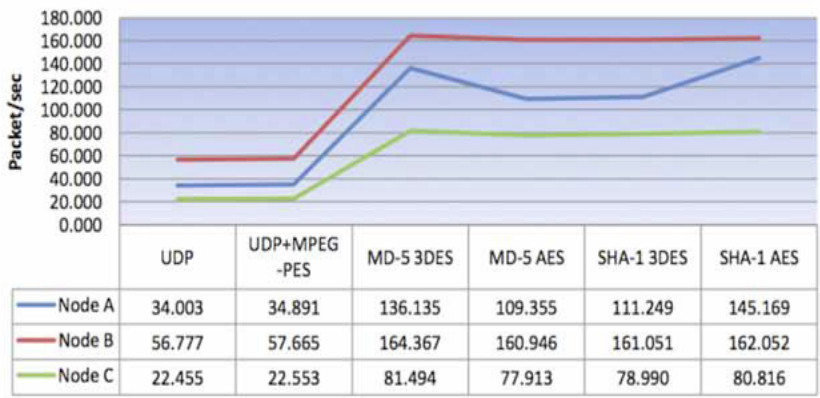

Figure 11. Packet data rate against different security schemes in a multi hop scenario.

## 4.5. Delay

We turn our attention to the data relating to the delay incurred by the security protocols. In each case, the data was gathered in the same manner to provide a set of balanced realistic results. Once the data had been collected from the testing period we needed to find the delay incurred by the security. From the data retrieved from the sending node (Node A), and in each instance of security protocol a subset of results was taken, which consisted of selecting twenty unique packets separated by a value of 10 packets. We then analysed these packets using WireShark, retrieving the Packet ID Number and the Time Stamp of each selected packet. Then the same process was applied to the data recorded on the receiving node (Node C (Multi-hop) and Node B (Single-hop)), except that the packets were inspected to find and match the Packet ID with the sending nodes twenty packets. By completing this process we now had a subset of twenty spaced out packets and the times that they were sent and received. This allowed us to subtract the sending time from the receiving time to produce the amount of time the packet was in transit also by taking the results for all the security combinations and comparing them we were able to see which combination was the most efficient for securing the network layer of MANETs. The results from calculating the delay incurred by adding the various security combinations are displayed in **Figure 12** and **Figure 13**.

**Figure 12** shows the average delay values, which have been calculated by adding up all the packet delay values and dividing them by the number of packets. As we can see from there is a definite trend to the amount of overhead added when each combination is applied. From this graph it is plain to see that AES incurs far less delay than 3DES. If we analyse **Figure 13**, which shows the plotted delay results for each packet of each cryptographic scheme we can see a smooth set of results in each case, with only some slight episodes relating to the SHA-1 combinations around packet 17, but the trend is immediately recovered at packet 18.

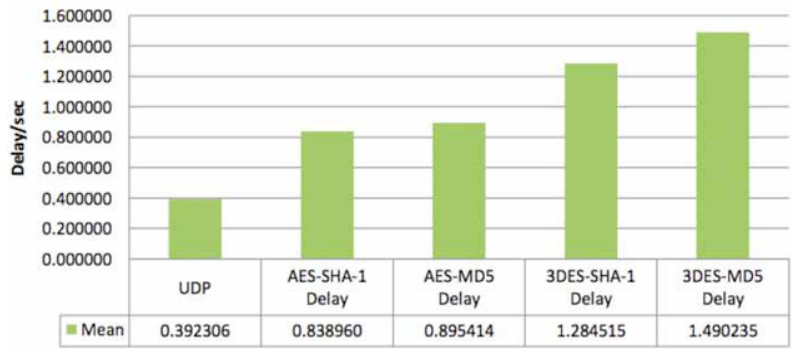

Figure 12. Average delay values of the different cryptographic schemes.

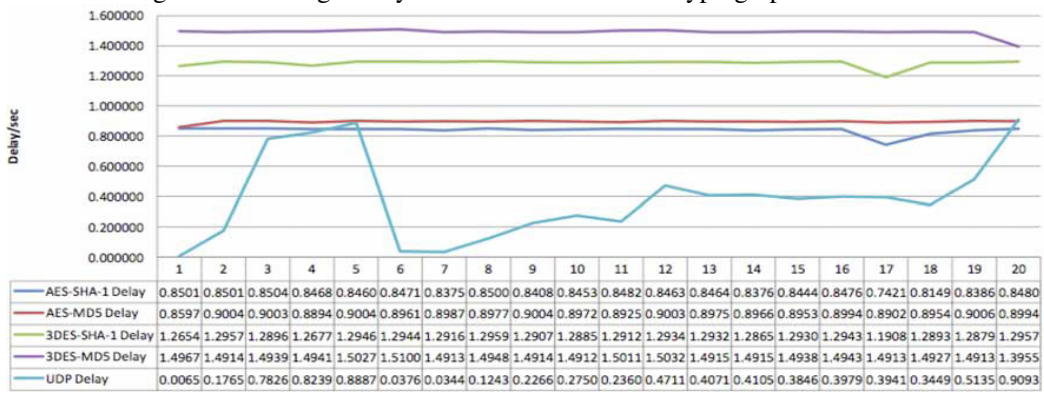

Figure 13. Plotted delay results for each packet of the different cryptographic schemes.

## 5. CONCLUSIONS AND FUTURE WORK

In this article we have assessed different cryptographic approaches to securing the OLSR protocol in MANETs using the *WMN test-bed*. We have additionally discussed the most important issues about security in MANETs and the most critical security benefits of the hybrid model of IPSec in MANETs. Furthermore, we have described the *WMN test-bed* software and the security implementation for MANETs. In closing we have evaluated the *WMN test-bed* results in terms of the time and space overhead that each mechanism introduces to MANETs. Our findings clearly show that the most efficient combination of algorithms used for authentication and encryption are SHA-1 and AES respectively. Using this combination over their counterparts will lead to a considerable reduction in processing time and delay on the network creating an efficient transaction moving towards satisfying resource constraints and security requirements. It would be the intention of future work to extend the number of devices in the test-bed scenario to find further performance data with the possibility of comparing previously untested algorithmic combinations in a MANET environment.

## ACKNOWLEDGEMENTS

This work was undertaken in the context of the project ICT- SEC-2007 PEACE (IP-based Emergency Applications and serviCes for nExt generation networks) with contract number 225654. The project has received research funding from the European 7th Framework Programme. The authors would like to thank the support of the ICT European Research Programme and all the partners in the PEACE project: PDM&FC, Insti- tuto de Telecomunicaciones, FhG Fokus, Thales, Telefonica I+D, CeBit.

**Authors**

**Emmanouil A. Panaousis** is currently a research Ph.D. student at Kingston University, UK, Faculty of Computing, Information Systems and Mathematics (CISM). He works within a team on Wireless Multimedia & Networking (WMN) Research Group. Emmanouil received his M.Sc. in Computer Science with distinction at the Department of Informatics of the Athens University of Economics and Business and his B.Sc. in Informatics and Telecommunications at the National and Kapodistrian University of Athens. Emmanouil is a student member of the British Computer Society, the IEEE and the IEEE Communications Society.

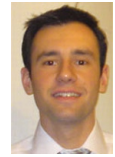

**George L M Drew** has recently graduated from Kingston University, UK, Faculty of Computing, Information Systems and Mathematics (CISM), with a 'First Class' BSc (Hons) degree in Computer Science and is currently considering an MSc in 'Wireless Communications'. While at Kingston University he worked with a team from the Wireless Multimedia Networking (WMN) Research Group.

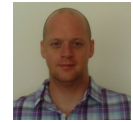

**Grant P. Millar** is currently a Ph.D student at the Faculty of Computing, Information Systems and Mathematics (CISM), Kingston University, London, UK. He works within a team on Wireless Multimedia Networking (WMN) as well as giving the occasional lecture on various subjects within the networking realm within the CISM faculty. Grant received his M.Sc in Networking and Data Communications at the faculty of CISM, Kingston University and his B.Sc in Computer Animation at the University of Portsmouth, UK. Grant is a student member of the IEEE and the IEEE Communications Society.

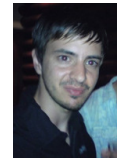

**Tipu Arvind Ramrekha** is a PhD student at Kingston University, UK, Faculty of Computing, Information Systems and Mathematics (CISM). There he is part of a research team on Wireless Multimedia Networking (WMN) and currently researching on, "Quality of Service routing for multimedia communications in Mobile Ad-hoc Networks for extreme emergency cases", in the CISM faculty. He holds a first class B. Eng. in Computer Engineering from NSIT, Delhi University, India. He then pursued successfully his MSc in Networking and Data Communications with management studies at Kingston University, London, UK where he received his MSc with a distinction and was awarded the best CISM postgraduate student and project awards. He is a member of the IEEE.

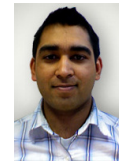

**Christos Politis** is a Reader (Assoc. Prof.) of Wireless Communications at Kingston University London, UK, Faculty of Computing, Information Systems and Mathematics (CISM). There he leads a research team on Wireless Multimedia & Networking (WMN) and teaches modules related to communications. Christos is the Field Leader for the 'Wireless Communications' and 'Networks and Data Communications' postgraduate courses at Kingston. Prior to this post, he was the Research and Development (R&D) project manager at Ofcom, the UK Regulator and Competition Authority. There he managed a number of projects across a wide range of technical areas including cognitive radio, polite protocols, radar, LE Applications, fixed wireless and mobile technologies. Christos' previous positions include telecommunications engineer with Intracom Telecom in Athens and for many years he was a post-doc research fellow in the Centre for Communication Systems Research (CCSR) at the University of Surrey, UK. He is being active with European research since 2000 and has participated in several EU, national and international projects. Christos was the initiator and the project manager of the IST UNITE project. He is a patent holder, and has published more than 80 papers in international journals and conferences and chapters in two books. Christos was born in Athens, Greece and holds a PhD and MSc from the University of Surrey, UK and a B.Eng. from the Technical University of Athens, Greece. He is a member of the IEEE and Technical Chamber of Greece.

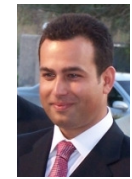